\documentclass{PoS}

\usepackage{lipsum}
\usepackage{lineno}
\usepackage{amsmath}
\usepackage{multicol}
\usepackage{mathtools}
\usepackage{cuted}
\usepackage{flushend}
\usepackage{amsmath}
\usepackage{graphicx}
\usepackage{bm}		% enables bold italic in math mode
\usepackage{multirow}
\usepackage{enumitem}
\usepackage{graphicx}
\newcommand\varpm{\mathbin{\vcenter{\hbox{%
  \oalign{\hfil$\scriptstyle+$\hfil\cr
          \noalign{\kern-.3ex}
          $\scriptscriptstyle({-})$\cr}%
}}}}
\newcommand\varmp{\mathbin{\vcenter{\hbox{%
  \oalign{$\scriptstyle({+})$\cr
          \noalign{\kern-.3ex}
          \hfil$\scriptscriptstyle-$\hfil\cr}%
}}}}

\title{Transversity and $\Lambda$ polarization at COMPASS}

\ShortTitle{Transversity and $\Lambda$ polarization at COMPASS}

\author{Andrea Moretti\\
        \rm{on behalf of the COMPASS Collaboration}\\
        \it{University of Trieste and INFN Trieste Section} \\
        E-mail: \email{andrea.moretti@ts.infn.it}}

\abstract{The study of $\Lambda$ polarization in Semi-Inclusive Deep Inelastic Scattering (SIDIS) with a transversely polarized target has often been suggested as a privileged way to access transversity, as the initial quark polarization can be transmitted to the hyperon and then observed via its weak decay. COMPASS measured the $\Lambda$ \textit{transversity-induced} polarization as a function of Bjorken $x$, of the fraction $z$ of the longitudinal momentum of the fragmenting quark carried by the $\Lambda$ and of the $\Lambda$ transverse momentum $P_T$. The results are compatible with zero in all regions of phase space. Under the hypothesis that transversity is a valence object, the previous knowledge of the transversity function allows to get information on the ratio of polarized and unpolarized $z-$integrated $\Lambda$ fragmentation functions. On the other hand, if the $\Lambda$ polarization is entirely due to the $s$ quark, or if a quark-diquark model is applied, information can be obtained on the $s$ quark transversity function.}

\FullConference{23rd International Spin Physics Symposium - SPIN2018 -\\
		10-14 September, 2018\\
		Ferrara, Italy}

\begin{document}

\section{Introduction}

The transverse polarization of neutral $\Lambda$ hyperons produced in SIDIS off transversely polarized nucleons was often indicated in the past as a promising channel to access transversity \cite{Baldracchini:1980uq,Artru:1990wq,Kunne:1993nq,Anselmino:2003wu,Avakian:2016rst}. If transversity is different from zero, the polarization of the fragmenting quark can in principle be transmitted to the final state hadron. Among all hadrons, $\Lambda$ hyperons are the most suited to polarimetry studies due to their self-analyzing weak decay, that allows to reconstruct the polarization $P_\Lambda$ through an angular asymmetry of the decay products:

\begin{equation}
    \frac{dN}{d\cos\theta} \propto 1 + \alpha P_\Lambda \cos\theta
\end{equation}
where $\alpha=0.642\pm0.013$ is the weak decay constant and $\theta$ is the proton emission angle with respect to the quantization axis, as calculated in the $\Lambda$ rest frame \cite{Commins:1983ns}. For $\bar{\Lambda}$s, the same relation holds for the antiproton, but with opposite sign of $\alpha$. In the current fragmentation region, at leading twist and in collinear approximation, the measured polarization for direct production $P_\Lambda^{exp}$ with respect to the outgoing quark spin can be written as \cite{Artru:1990wq,Mulders:1995dh,Barone:2003fy}:

\begin{equation}
    P^{exp}_\Lambda(x,z) = f p_t D_{NN}  \frac{\sum_{q(\bar{q})}e_q^2h_1^q(x)H_1^{\Lambda/q}(z)}{\sum_{q(\bar{q})}e_q^2f_1^q(x)D_1^{\Lambda/q}(z)}
\end{equation}
where $f$ is the dilution factor, $p_t$ the target polarization and $D_{NN} = \frac{2(1-y)}{1+(1-y)^2}$ is the QED depolarization factor, being $y$ the fractional energy of the virtual photon. If the intrinsic transverse momentum $k_T$ of the initial quark is taken into account, the expression for $P_\Lambda$ involves more terms \cite{Boer:1999uu}. 

Studies on $\Lambda$ \textit{transversity-induced} polarization have been performed in COMPASS using data collected in 2002-2004 with a transversely polarized deuteron target. However, due to limitations in statistics and spectrometer acceptance, and to the lack of particle identification for large part of the data set, the results of that analysis \cite{Ferrero} are not considered now. The analysis presented here supersedes a previous study based on 2007 data only \cite{Negrini:2009zz}. It has been performed using the COMPASS data collected both in 2007 and 2010 with a 160 GeV/$c$ longitudinally polarized muon beam from the CERN SPS and a transversely polarized NH$_3$ target with proton polarization $\langle p_t \rangle \approx 0.8$ and dilution factor $\langle f \rangle \approx 0.15$. 

\vspace{0.5cm} \ \\ 
\section{Results}
\subsection{Data selection}
 
DIS events have been selected asking for the photon virtuality $Q^2>1$ (GeV/$c$)$^2$, Bjorken variable $x>0.003$, $0.1<y<0.9$ and final state hadronic mass $W>$ 5 GeV/$c^2$. The reconstruction of $\Lambda$s and $\bar{\Lambda}$s relied on the identification of secondary vertices $V^0$ with two outgoing oppositely charged tracks ($\Lambda \rightarrow p \pi^-$, $\bar{\Lambda} \rightarrow \bar{p} \pi^+$; branching ratio $BR=63.9\%$), well separated from the primary vertex, which includes the incoming and outgoing muons. The long lifetime ($\tau = 2.632 \pm 0.020 \cdot 10^{-10}$ s) ensures a good separation between primary and secondary vertices. Momenta higher than 1 GeV/$c$ have been required for the outgoing particles to ensure a good quality of the reconstructed tracks. Also, a limit has been set on the angle $\theta_{coll}$ between the reconstructed line of flight $\vec{p}_{V^0}$ and the vector linking primary and secondary vertex $\vec{v}$:

\begin{equation}
\theta_{coll} = \mathrm{arccos}\left( \frac{\vec{p}_{V^0}\cdot \vec{v}}{|\vec{p}_{V^0}||\vec{v}|}\right) < \mathrm{7 \ mrad} .
\end{equation}

To suppress the contribution to the $V^0$ sample coming from photon conversion into $e^+e^-$ pairs, the transverse momentum $p_T$ of each decay particle with respect to the $V^0$ line of flight had to be larger than 23 MeV/$c$. The RICH information has been used to reject all $\Lambda$ ($\bar{\Lambda}$) candidates with an identified $e^+$, $\pi^+$ or $K^+$ (respectively $e^-$, $\pi^-$ and $K^-$). This vetoing allowed to maximize the $\Lambda$ ($\bar{\Lambda}$) sample while keeping the signal-over-background ratio relatively low. %PID was based on the calculation of the maximum likelihood $\mathcal{L}$ for five mass hypotheses ($e$, $\mu$ $K$, $\pi$, $p$) and for the background for a given number of collected Cherenkov photons. To attribute a mass hypothesis $M$ to a particle, $\mathcal{L}_M$ was requested to be the highest and its ratio to the background hypothesis to be larger than a certain factor, tipically ranging from 1.9  to 3.0. This approach has been successfully applied to particles with momentum up to than 50 GeV/$c$, value at which the pion/kaon separation becomes difficult. Beyond this limit, the highest likelihood was asked not to be the one associated to the pion or kaon mass. 

\begin{center}
\begin{figure}[!h]
    \centering
    \includegraphics[width=0.5\textwidth]{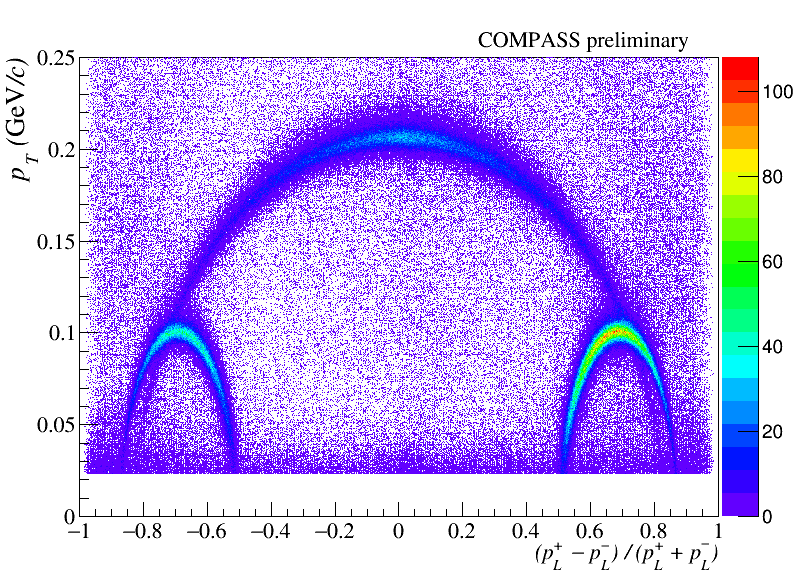}
    \caption{Armenteros plot after all but mass cut. }
        \label{fig:armenteros}

\end{figure}
\end{center}

\begin{center}
\begin{figure}[!h]
    \centering
    \includegraphics[width=0.45\textwidth]{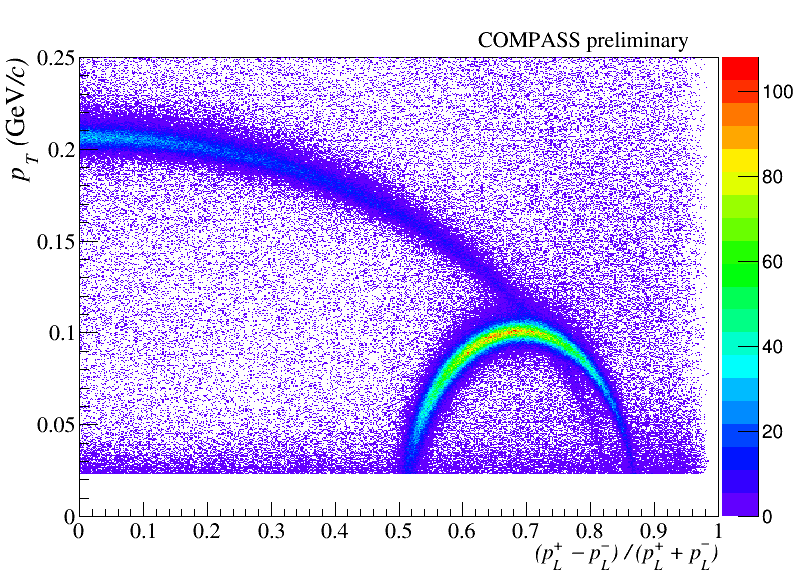}
    \includegraphics[width=0.45\textwidth]{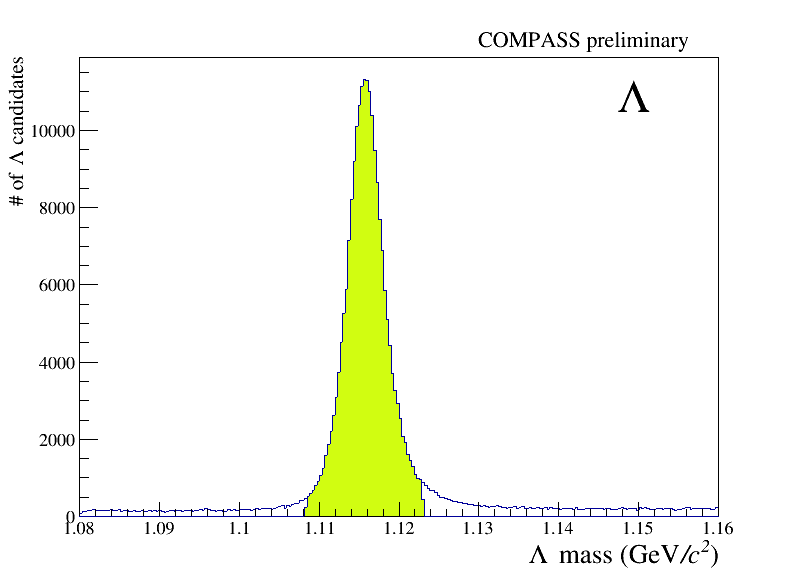}
    \includegraphics[width=0.45\textwidth]{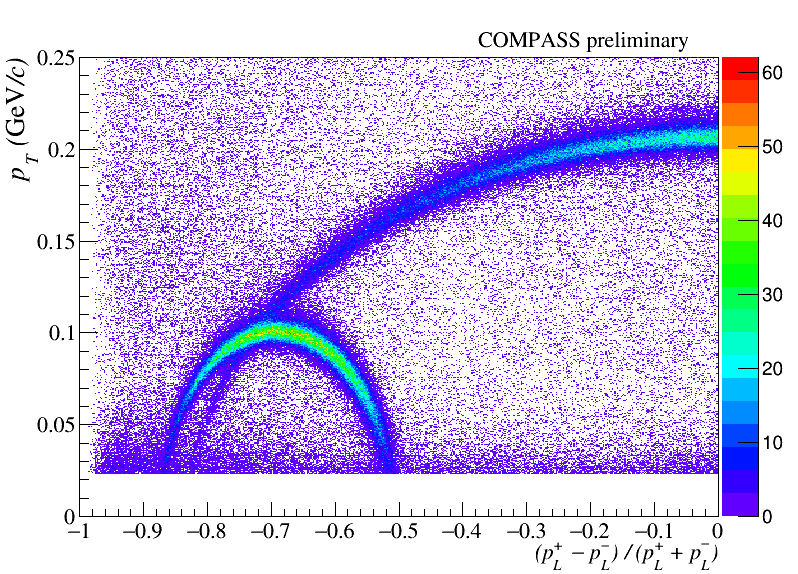}
    \includegraphics[width=0.45\textwidth]{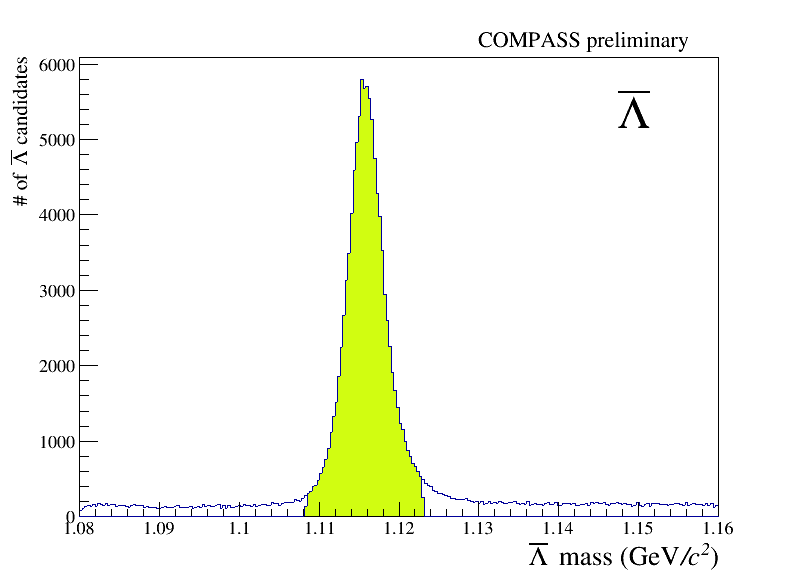}
    \caption{Armenteros plot after all but mass cut for positive and negative longitudinal momentum asymmetry, together with the corresponding invariant masses. Filled areas indicate candidates inside the 3$\sigma$ cuts.}
    \label{fig:armenteroslr}

\end{figure}
\end{center}

The Armenteros plot \cite{Armenteros1, Armenteros2} for candidates surviving all the selection steps (except for the mass cut) is shown in Figure~\ref{fig:armenteros}. The $K_s^0$ contribution, still visible as the large, symmetric arc centered at zero, is further reduced with a cut on the invariant mass of the pair. A selection of the left and right part of the Armenteros plot (Figure~\ref{fig:armenteroslr}), based on the sign of the longitudinal momentum asymmetry $\frac{p_L^+-p_L^-}{p_L^++p_L^-}$, allows to separate $\bar{\Lambda}$s (on the left) from $\Lambda$s (on the right). The quantities $p_L^+$ and $p_L^-$ indicate the longitudinal momentum with respect to the $V^0$ line of flight of the positive and negative decay particle respectively. In Figure~\ref{fig:armenteroslr} the $\Lambda$ and $\bar{\Lambda}$ invariant mass spectra, before the mass cut, are given together with the corresponding section of the Armenteros plot. Depending on the kinematic bin, the signal-over-background ratio ranges from 5.7 to 54.9. The invariant mass spectra have been fitted with a superposition of a gaussian function and a constant term. Background has been evaluated with the sideband method considering two equally wide intervals on the left and on the right of the mass peak. Candidates have been selected within a $\pm 3\sigma$ range, where $\sigma$ = 2.45 MeV/$c^2$. The total statistics after background subtraction is given in Table~\ref{table:stat}.

A significant fraction of $\Lambda$ particles originates from the decay of heavier hyperons. It has been estimated with LEPTO \cite{Adolph:2013dhv} that in the COMPASS regime 63\% of the produced $\Lambda$s and 68\% of $\bar{\Lambda}$s originate from direct string fragmentation. In this analysis candidates coming from indirect production contribute to the measured polarization, but no attempt has been made to estimate their contribution to the systematic error.
 
\begin{center}
\begin{table}[!h]
    \centering
    \begin{tabular}{ c c c  }
 year & $\Lambda$ & $\bar{\Lambda}$\\ \hline 
 2007 & 95 125  $\pm$ \ 315 & 44 911  $\pm$ \ 227\\  
 2010 & 201 421  $\pm$ \ 466 & 99 552  $\pm$ \ 336 \\ \hline
 total & 296 546  $\pm$ \ 562 &  144 463  $\pm$ \ 405\\ \hline
\end{tabular}
    \caption{Partial and total statistics for $\Lambda$ and $\bar{\Lambda}$ candidates after background subtraction.}
\label{table:stat}
\end{table}

\end{center}

\bigskip
 
\subsection{Extraction of $\Lambda$ ($\bar{\Lambda}$) polarization}
The data have been collected by using simultaneously three target cells with the first and the last cells (each half of the length of the central cell) oppositely polarized with respect to the central one. By suitably combining the data, instrumental asymmetries could be limited to negligible values. The principle of the measurement and the data analysis are already described in several publications \cite{Alekseev:2010rw,Adolph:2012sn} and will not be repeated here. 

We measured the $\Lambda$ polarization along the spin of the fragmenting quark, the determination of the quantization axis being done on a event-by-event basis. The initial quark spin, assumed to be aligned to the nucleon spin, is vertical in the laboratory frame. Its transverse part has azimuthal angle in the gamma-nucleon system $\phi_S$ and zero polar angle. The spin of the quark after the interaction with the virtual photon is reflected with respect to the normal to the scattering plane \cite{Barone:2003fy}:  $\phi_S^\prime=\phi_S+\pi$. For direct production, the azimuthal angle of the spin of the final state hyperon is assumed aligned with $\phi_S^\prime$. The number of $\Lambda$s emitting a proton in a given $\cos\theta$ range as measured in the $\Lambda$ rest frame, from a given target cell with a given direction of the target polarization can be expressed as:

\begin{equation}\label{N}
    \mathcal{N}_i^{(\prime)} = \Phi_i^{(\prime)} \ \rho_i^{(\prime)} \ \bar{\sigma} \ (1 \varpm \alpha P_\Lambda^{exp} \cos\theta) \ A_i^{(\prime)}(\cos\theta) 
\end{equation}
with $i=1,2$ indicating either the central or the outer cells respectively, $\Phi_i^{(\prime)}$ the muon flux, $\rho_i^{(\prime)}$ the number of nucleons per cm$^2$ in the target, $\bar{\sigma}$ the cross section for the production of $\Lambda$s and $A_i^{(\prime)}(\cos\theta)$ the acceptance term, which includes both geometrical acceptance and spectrometer efficiency. Primed quantities refer to subperiods taken after polarization reversal.  After background subtraction, the four quantities $\mathcal{N} $ of Equation~\ref{N} are combined into a single quantity, the \textit{double ratio} $ \varepsilon = \frac{\mathcal{N}_1 \mathcal{N}_2'}{\mathcal{N}_1'\mathcal{N}_2}$. The flux cancels in the double ratio; acceptances also cancel, under the hypothesis that the ratio of acceptances between the different target cells remains constant, for small variations of the apparatus performances (the so-called "reasonable assumption" \cite{Ageev:2006da}). As a result, for small values of $\alpha P_\Lambda$, we can write: 

\begin{equation}
    \varepsilon = \frac{\mathcal{N}_1 \mathcal{N}_2'}{\mathcal{N}_1'\mathcal{N}_2} \approx 1+4\alpha P_\Lambda^{exp}\cos\theta.
\end{equation}

For each kinematic bin in $x$, $z$ and $P_T$ the data sample was divided into eight $\cos\theta$ bins and $P_\Lambda^{exp}$ extracted with a linear fit of the quantities $\varepsilon_j$, with $j=1,\dots,8$. The $\Lambda$ and $\bar{\Lambda}$ polarizations have been measured in the following cases:

\begin{itemize}[itemsep=0pt]
    \item all candidates considered;
    \item high $z$ region: $z>0.2$ and $x_F>0$, where $x_F$ is the Feynman scaling variable;
    \item low $z$ region: $z<0.2$ or $x_F<0$;
    \item high $x$ region: $x>0.032$, where transversity for $u$ and $d$ quarks is known to be different from zero;
    \item low $x$ region: $x<0.032$;
    \item high $P_T$ region: $P_T>0.5$ GeV/$c$;
    \item low $P_T$ region: $P_T<0.5$ GeV/$c$.
\end{itemize}

\begin{center}
\begin{figure}[h!]
    \centering
    \includegraphics[width=0.45\textwidth]{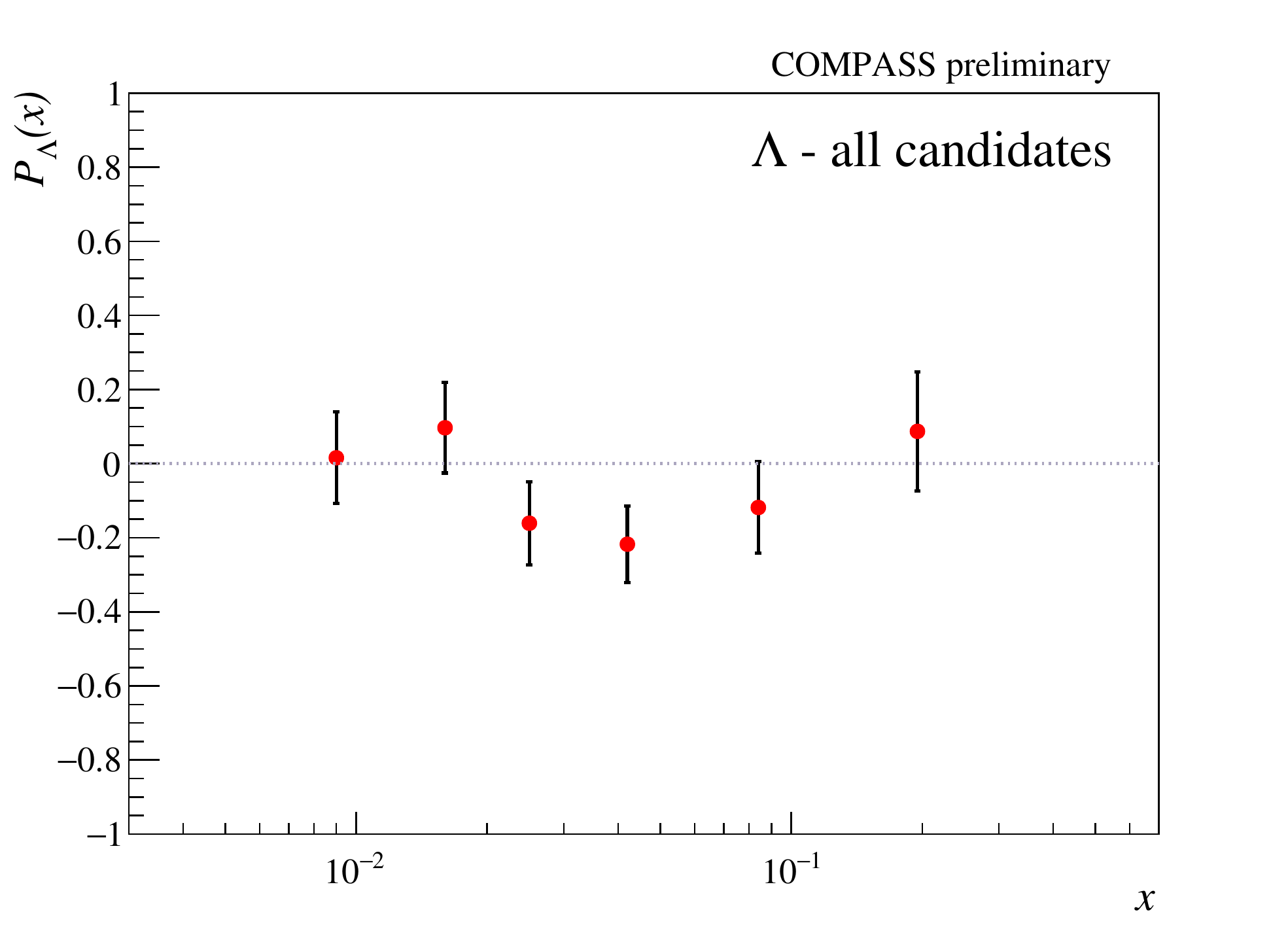}
    \includegraphics[width=0.45\textwidth]{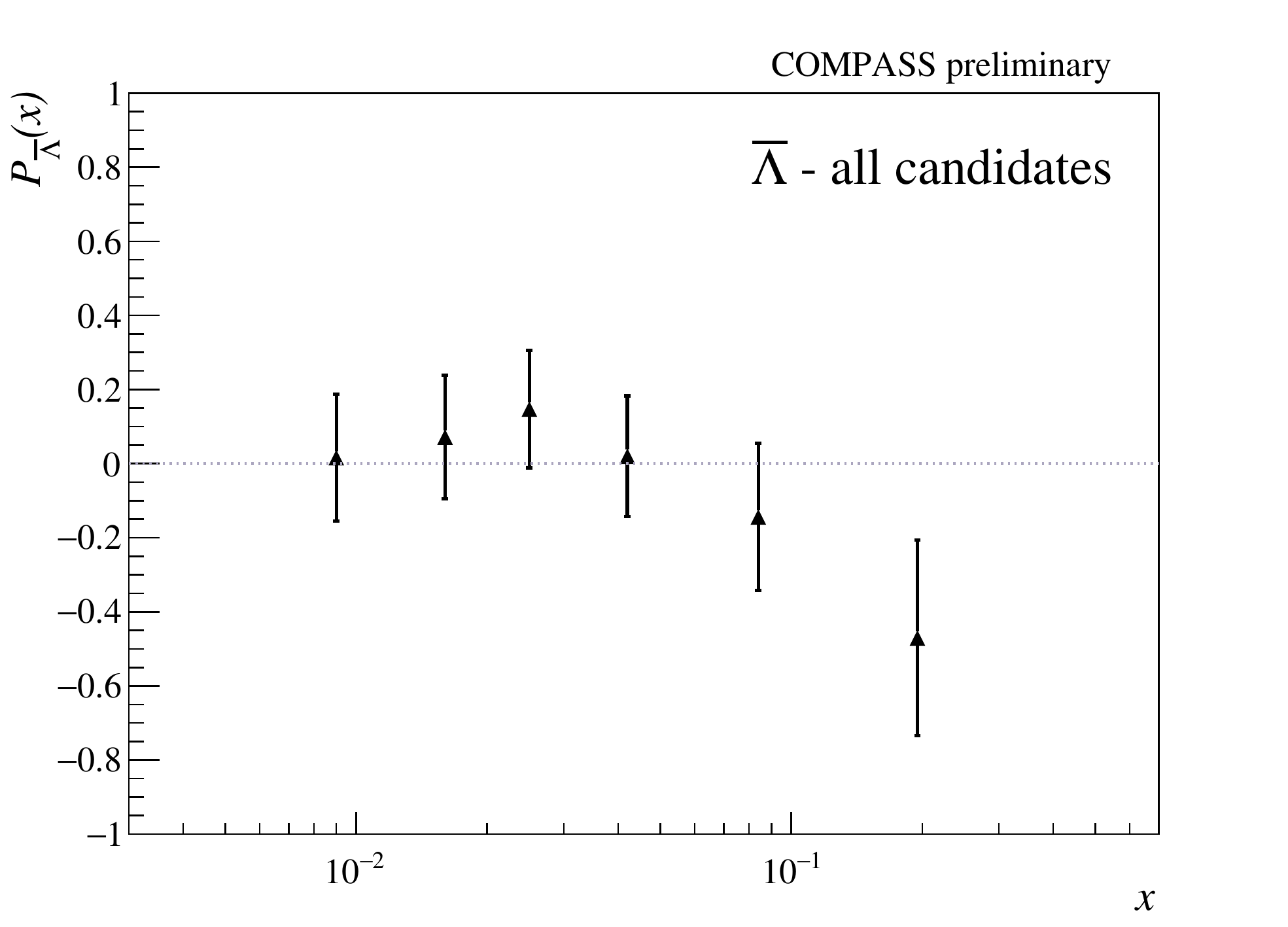}\\
    \includegraphics[width=0.45\textwidth]{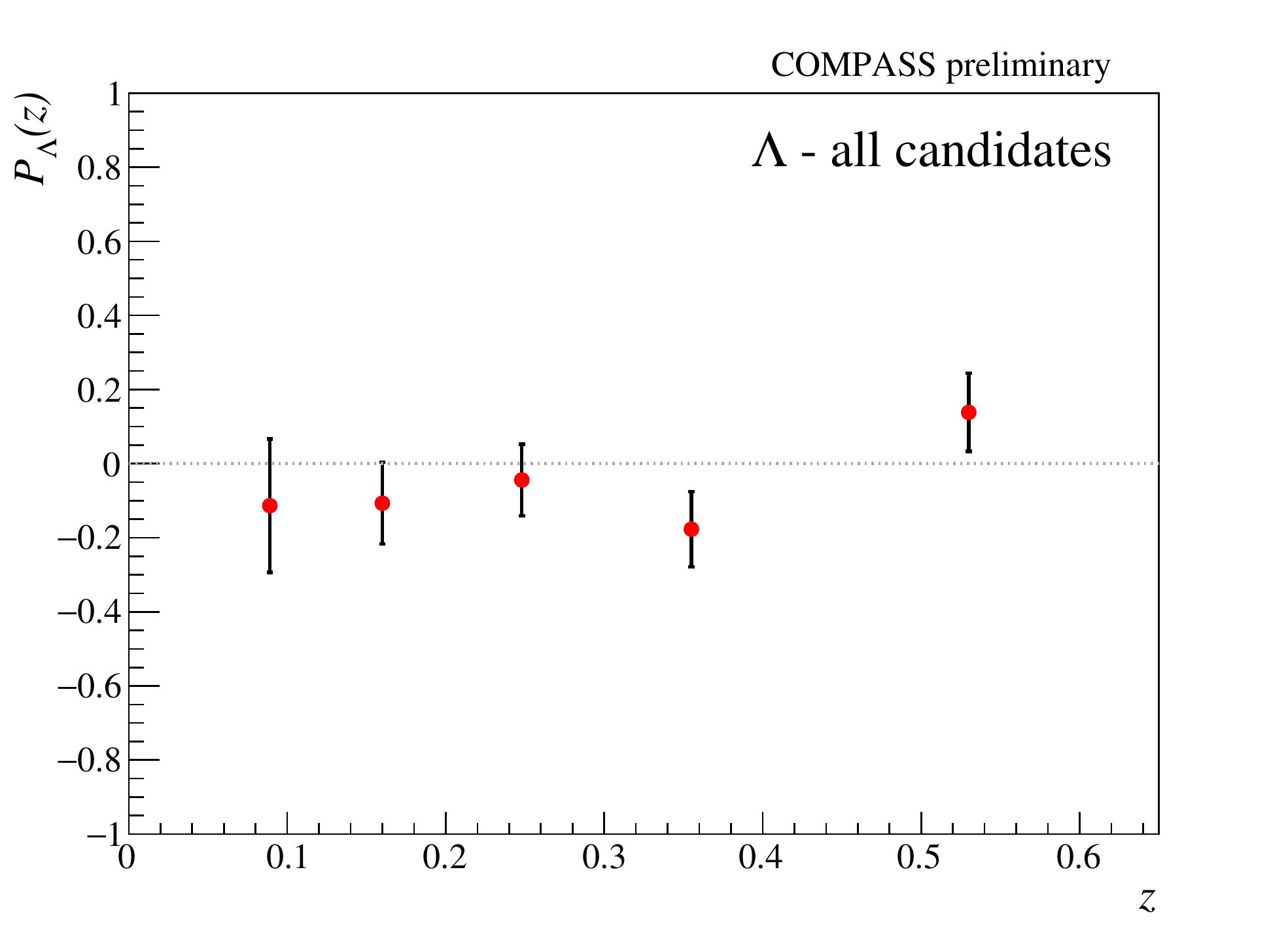}
    \includegraphics[width=0.45\textwidth]{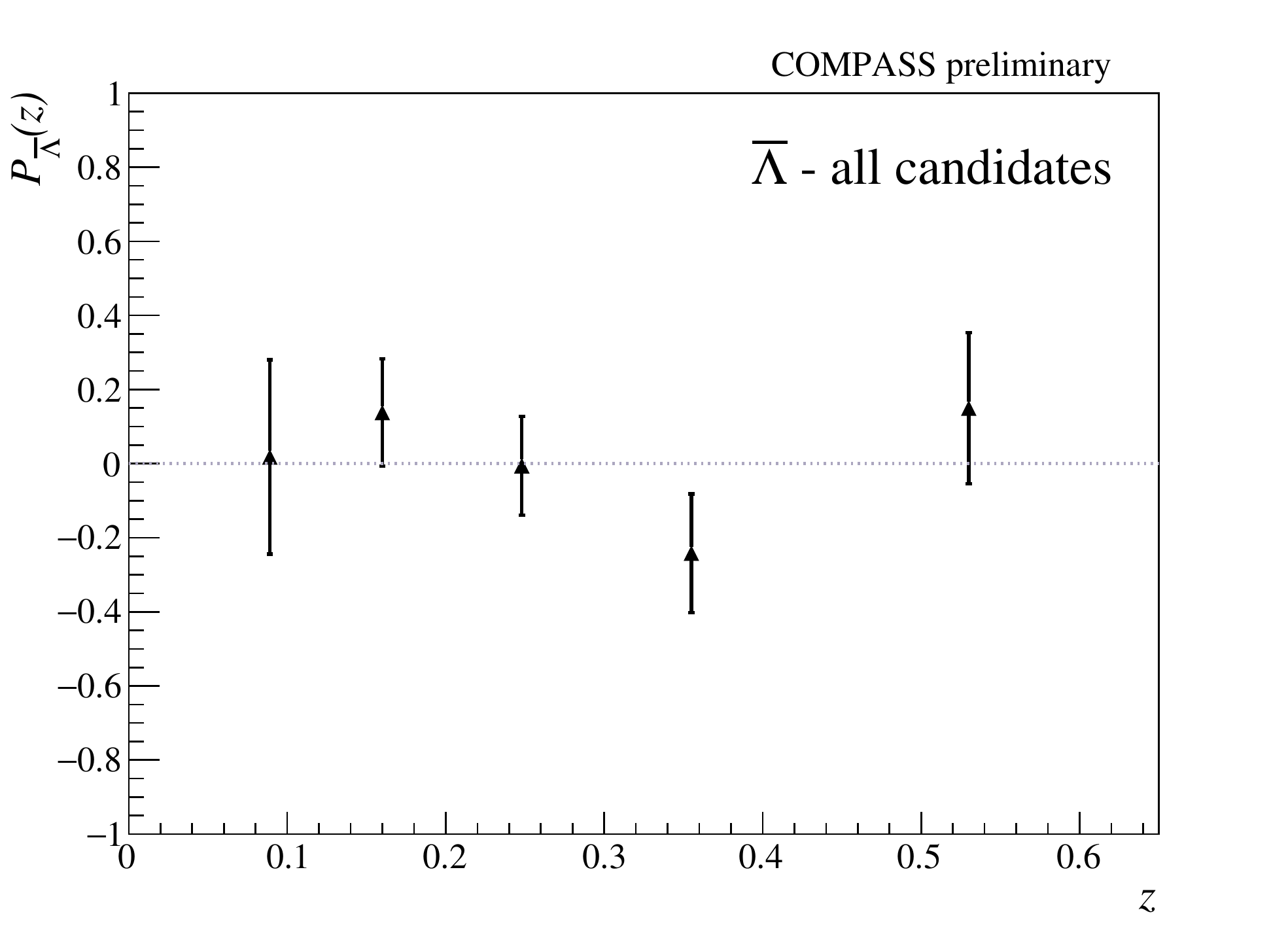}\\
    \includegraphics[width=0.45\textwidth]{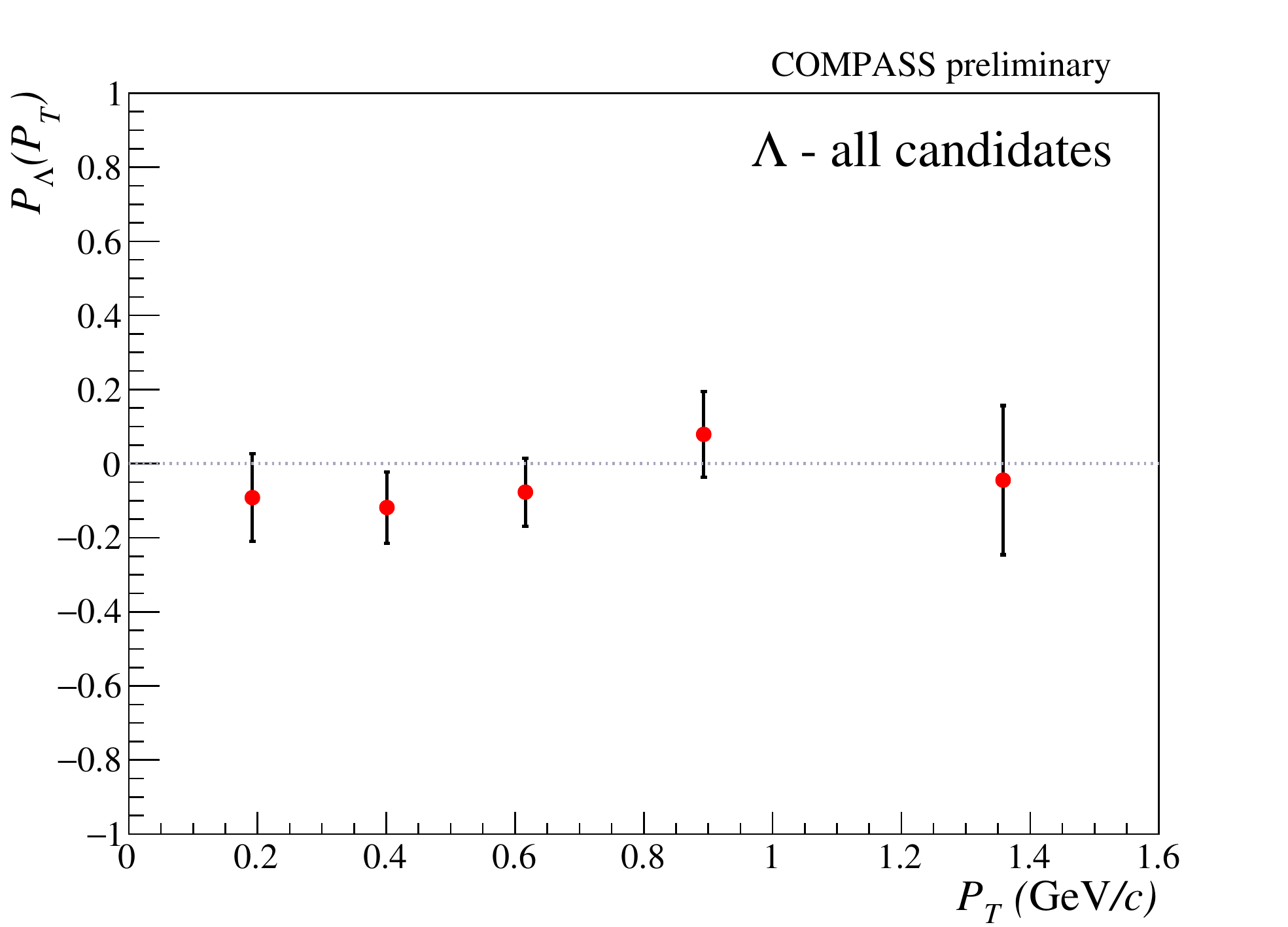}
    \includegraphics[width=0.45\textwidth]{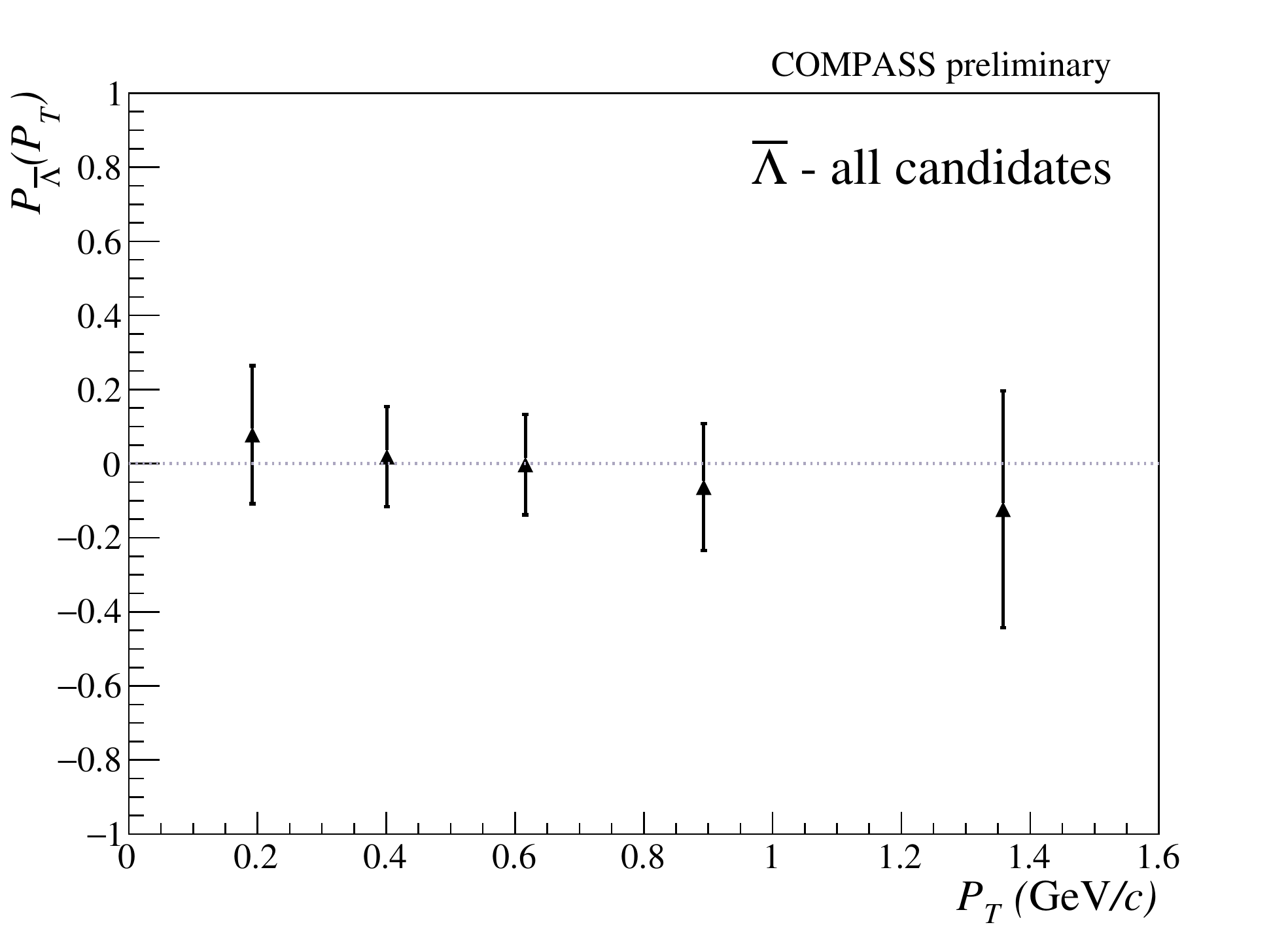}
    \caption{Polarizations for the "all candidates" case. Uncertainties are statistical only.}
    \label{fig:all_pol}
    
    \vspace{0.5cm} 
    \includegraphics[width=0.45\textwidth]{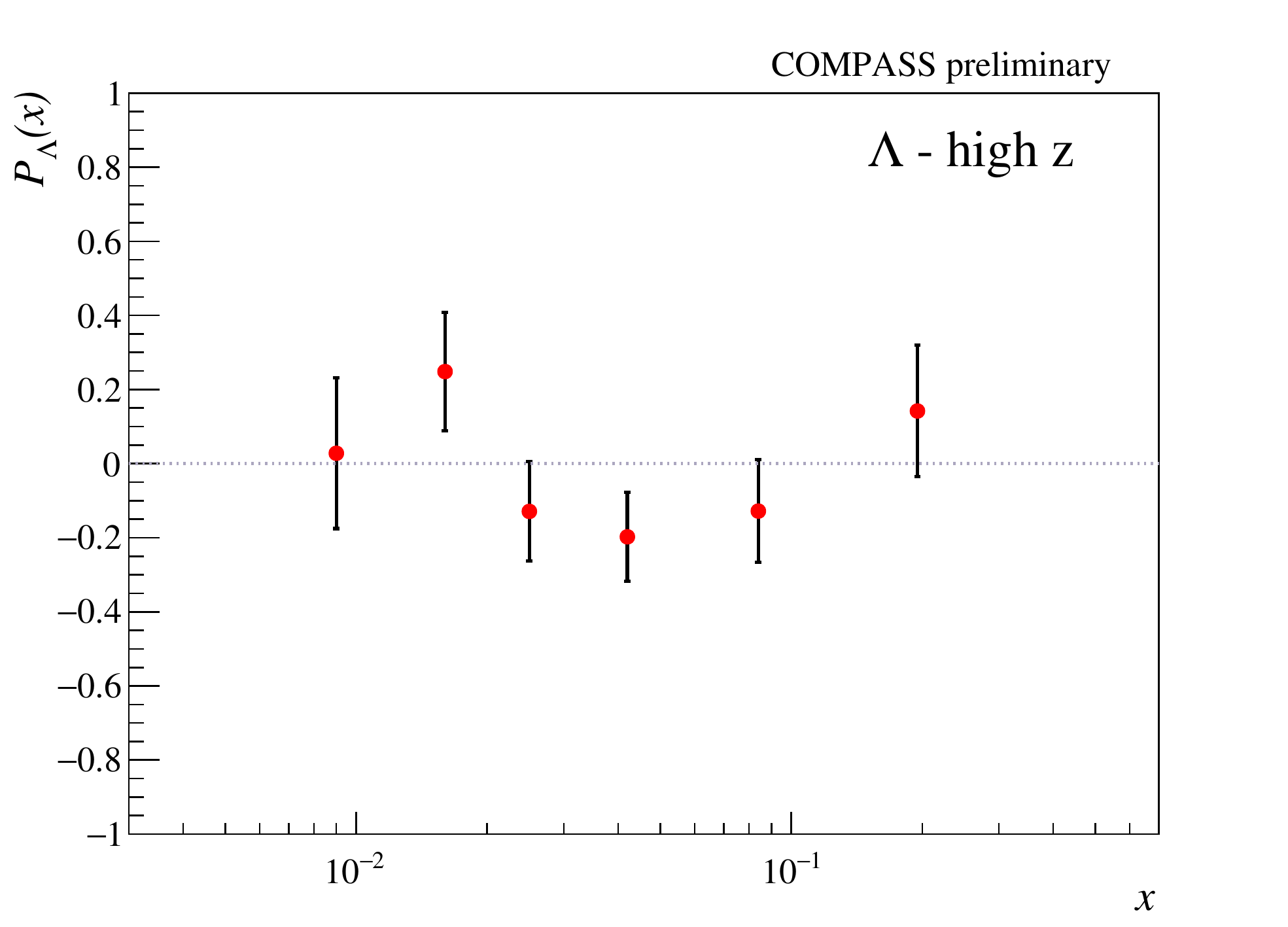}
    \includegraphics[width=0.45\textwidth]{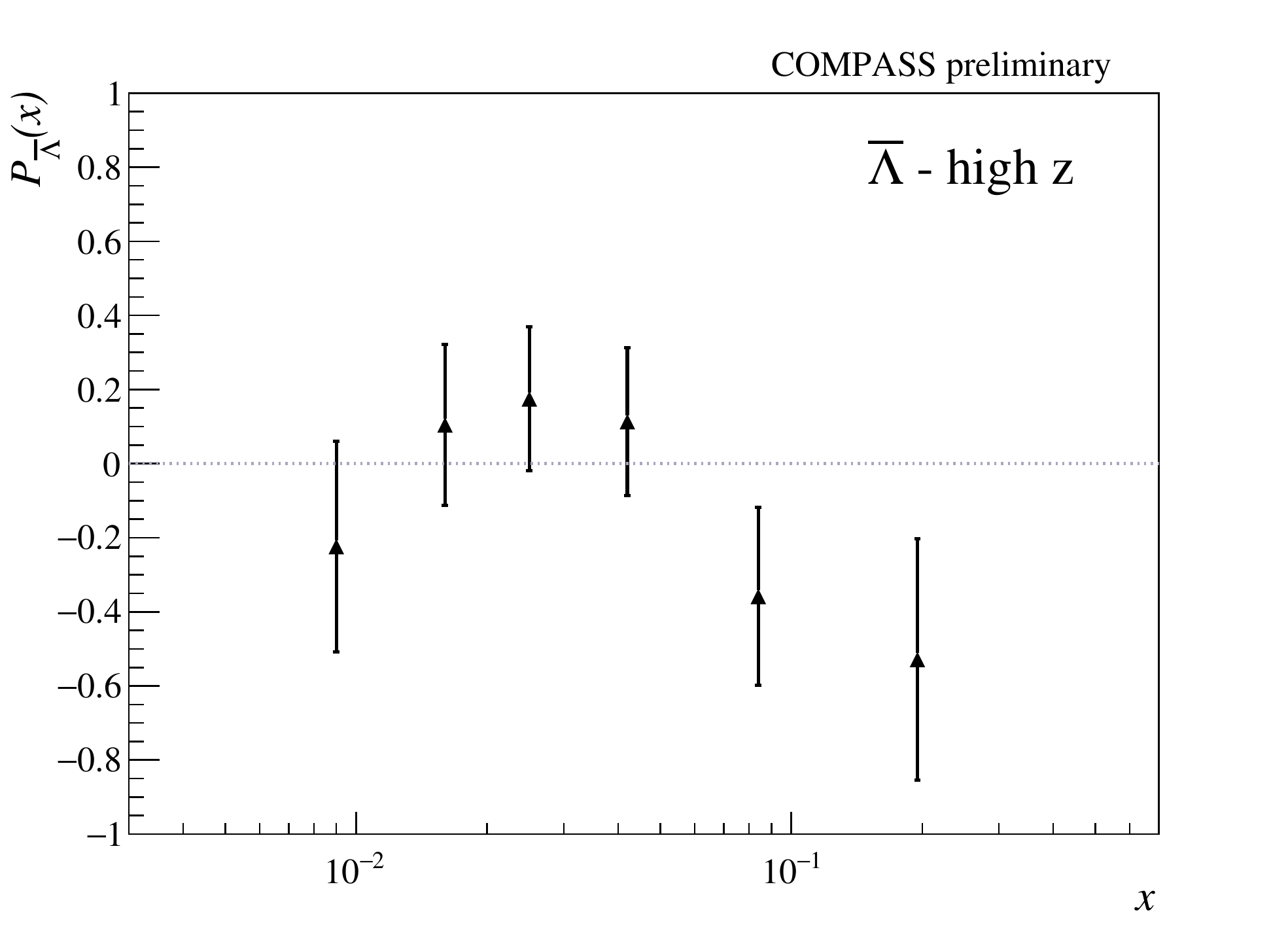}
    \caption{Polarizations for the "high $z$" case. Uncertainties are statistical only.}
    \label{fig:cfr_pol}
\end{figure}
\end{center}

Polarizations have been found compatible with zero within the experimental errors in all regions. Polarizations for the "all candidates" case are given, divided by $f p_T D_{NN}$, in Figure~\ref{fig:all_pol}. Polarization as function of $x$ is presented for the high-$z$ region in Figure~\ref{fig:cfr_pol}. 

Systematic errors are both multiplicative and additive. In the former category, the main contribution comes from the uncertainty on the dilution factor $f$; the uncertainty on mean values of the target polarization, almost negligible for 2010 data only, is on average (2007+2010) smaller than 1$\%$, while the relative uncertainty on the weak decay constant $\alpha$ is 2$\%$. The uncertainty on the depolarization factor $ D_{NN}$ is negligible. Several tests have been performed to estimate the additive contribution to systematic errors, namely period compatibility, $K^0_s$ polarization and false polarization. The overall systematic error is estimated to be 80\% of the statistical one.

\vspace{-1cm} 
\section{Interpretation}
The transversity-induced polarization of $\Lambda$ hyperons in the current fragmentation region can be written as:

\begin{equation}\label{pol}
    P_\Lambda(x,z) = \frac{\sum_{q(\bar{q})}e_q^2 h_1^{q(\bar{q})}(x) H_1^{\Lambda/q(\bar{q})}(z)}{\sum_{q(\bar{q})}e_q^2 f_1^{q(\bar{q})}(x) D_1^{\Lambda/q(\bar{q})}(z)}.
\end{equation}

Values of $h_1^u$ and $h_1^d$, as extracted by a global analysis of SIDIS and $e^+e^-$ data, are sizably different from zero in the valence region, while for $\bar{u}$ and $\bar{d}$ they are compatible with zero \cite{Martin:2014wua}. Assuming that the contribution to $P_\Lambda$ from the $\bar{s}$ quark is also negligible, the numerator of Equation~\ref{pol} gets simplified in this way:

\begin{equation}
    \sum_{q(\bar{q})}e_q^2 h_1^{q(\bar{q})}(x) H_1^{\Lambda/q(\bar{q})}(z) \approx \sum_{q=uds}e_q^2 h_1^{q}(x) H_1^{\Lambda/q}(z).
\end{equation}

Neglecting the antiquarks contributions to the unpolarized $\Lambda$ fragmentation process, the denominator of Equation~\ref{pol} can be rewritten as:

\begin{equation}
        \sum_{q(\bar{q})}e_q^2 f_1^{q(\bar{q})}(x) D_1^{\Lambda/q(\bar{q})}(z) \approx \sum_{q=uds}e_q^2 f_1^{q}(x) D_1^{\Lambda/q}(z).
\end{equation}

Isospin symmetry fixes $D_1^{\Lambda/d} = D_1^{\Lambda/u}$ and $H_1^{\Lambda/d}= H_1^{\Lambda/u}$. As for the $s$ fragmentation functions, a naive approach suggests $D_1^{\Lambda/s} = c_1 D_1^{\Lambda/u}$, being $c_1$ a constant. The quantity $1/c_1$ is often referred to as \textit{strangeness suppression factor}; in \cite{Yang:2002gh} its value, obtained from a fit of experimental data for production of baryon in $e^+e^-$, is $\lambda_\Lambda=1/c_1 \approx 0.44$. With these simplifications, Equation~\ref{pol} turns into:

\begin{equation}
    P_{\Lambda}(x,z) = \frac{\left[4h_1^u(x)+h_1^d(x)\right] H_1^{\Lambda/u}(z) + h_1^s(x)H_1^{\Lambda/s}(z)}{\left[4f_1^u(x)+f_1^d(x)+c_1f_1^s(x)\right]D_1^{\Lambda/u}(z)}.
\end{equation}

Several scenarios have been suggested to interpret $\Lambda$ polarization (see e.g. \cite{Anselmino:2000ga}). Here we choose three different options:

\begin{itemize}[itemsep=0pt]
    \item transversity is a valence object;
    \item polarization is entirely carried by the $s$ quark;
    \item quark-diquark model holds.
\end{itemize}

We referred to the CTEQ5D library \cite{Buckley:2014ana} for the unpolarized parton distribution functions and to previous extractions of transversity for $u$ and $d$ quarks \cite{Martin:2014wua}, properly fitted and extrapolated to the $x$ and $Q^2$ values of interest for this analysis. The statistical uncertainty at the maximum of $h_1^u$ and $h_1^d$ amounts to less than 5\%; for a detailed study see \cite{Franco:fe18}.

\vspace{1cm}
- case 1: {\it{Transversity is a valence object} \\}
If transversity is a valence object, then $h_1^s \approx 0$ and the measurement of $P_{\Lambda}(x)$ allows to extract the ratio $\mathcal{R}$ of the $z-$integrated fragmentation function $H_1^{\Lambda/u} / D_1^{\Lambda/u}$:

\begin{equation}
\mathcal{R}(x)= \frac{\int dz H_1^{\Lambda/u}(z)}{\int dz D_1^{\Lambda/u}(z)} = \frac{4f_1^u(x)+f_1^d(x)+c_1f_1^s(x)}{4h_1^u(x)+h_1^d(x) } P_{\Lambda}(x). 
\end{equation}

In the following Table~\ref{table:R} the mean value of $\mathcal{R}$ is given for three values of $c_1$. The results, which show a weak dependence on $c_1$, are all negative and compatible with zero.

\begin{center}
\begin{table}[!h]
    \centering
    \begin{tabular}{ c c  }\hline
 $c_1$ & $\langle \mathcal{R} \rangle \pm \sigma_{\langle\mathcal{R}\rangle}$ \\ \hline 
 2 & -0.39 $\pm$ 0.73 \\  
 3 & -0.38 $\pm$ 0.75 \\   
 4 & -0.37 $\pm$ 0.76 \\ \hline
\end{tabular}
\caption{Mean value $\langle \mathcal{R} \rangle$ for three choices of $c_1$. Uncertainties are statistical only.}
\label{table:R}
\end{table}

\end{center}

\begin{center}
\begin{figure}[!h]
    \centering
    \includegraphics[width=0.5\textwidth]{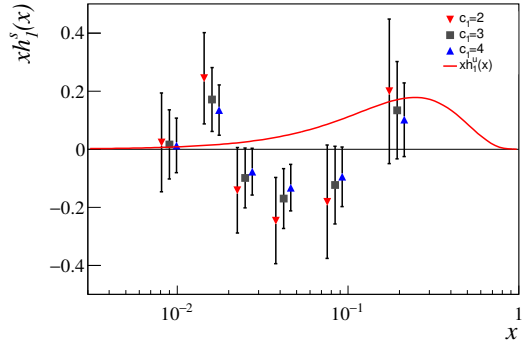}
    \caption{Extracted values of $xh_1^s(x)$ for the three options $c_1=2,3,4$. The $u$ quark transversity curve is drawn for comparison. Uncertainties are statistical only.}
        \label{fig:xh1s}

\end{figure}
    
\end{center}

- case 2: {\it{Polarization carried by $s$ quark only} \\}
If the transversity-induced polarization $P_{\Lambda}$ is entirely due to the $s$ quark, as from the SU(3) non relativistic quark model, $H_1^{\Lambda/u}$ can be neglected in favour of a $H_1^{\Lambda/s}$ contribution almost the same size of $D_1^{\Lambda/s}$, giving:

\begin{equation}
\begin{aligned}
        P_{\Lambda}(x,z) & = \frac{h_1^s(x)H_1^{\Lambda/s}(z)}{\left[4f_1^u(x)+f_1^d(x)+c_1f_1^s(x)\right]\frac{1}{c_1}D_1^{\Lambda/s}(z)} \\ 
            & \approx \frac{c_1 h_1^s(x)}{4f_1^u(x)+f_1^d(x)+c_1f_1^s(x)}.
\end{aligned}
\end{equation}
With a simple rearrangement of the terms, the transversity of the $s$ quark can be extracted. In Figure~\ref{fig:xh1s} the quantity $xh_1^s(x)$ is given for various $c_1$'s and compared to the fitted $xh_1^u(x)$ distribution.

\begin{center}
\begin{figure}[!h]
    \centering
    \includegraphics[width=0.5\textwidth]{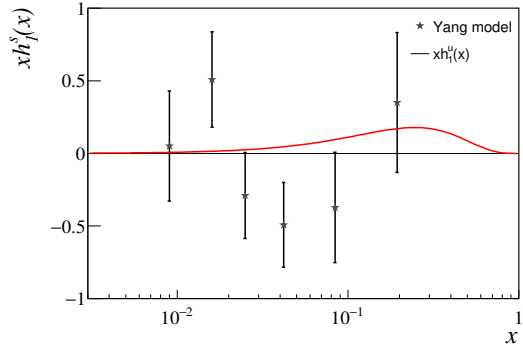}
    \caption{Extracted values of $xh_1^s(x)$ according to Yang's quark-diquark model. The $u$ quark transversity curve is drawn for comparison. Uncertainties are statistical only.}
    \label{fig:xh1s_y}

\end{figure}
    
\end{center}

\vspace{1cm} 
- case 3: {\it{Quark-diquark model} \\}
Within the framework of the Yang quark-diquark model \cite{Yang:2001sy, Jakob:1997wg}, the quark-to-$\Lambda$ unpolarized fragmentation function are written using $a_D^q(z)$ terms, where $D$  can be either a "scalar" ($S$) or a "vector" ($V$) according to the spin configuration of the emitted anti-diquark. Analogously, transversely polarized fragmentation functions are modeled via $\hat{a}_D^q(z)$ terms. Both $a_D^q(z)$ and $\hat{a}_D^q(z)$ terms enter into the definition of the calculable flavour structure functions $F^{(u/s)}$ and the spin structure functions $\hat{W}$. In this model the $\Lambda$ polarization can be written as:

\begin{center}
  \begin{equation}
        P_{\Lambda}(x,z) = \frac{\left(4h_1^u(x)+h_1^d(x)\right) \cdot \frac{1}{4}\left[ \hat{W}_S^{(u)}(z)F_S^{(u/s)}(z) - \hat{W}_V^{(u)}(z)F_M^{(u/s)}(z)\right] +  h_1^s(x)\hat{W}_S^{(s)}(z)}   {\left(4f_1^u(x)+f_1^d(x)\right) \cdot \frac{1}{4}\left[ F_S^{(u/s)}(z) +3 F_M^{(u/s)}(z)\right] +  f_1^s(x)}
\end{equation}
\end{center}
Here, $F_S^{(u/s)}(z) = a_S^{(u)}(z) / a_S^{(s)}(z)$, $F_M^{(u/s)}(z) = a_V^{(u)}(z) / a_S^{(s)}(z)$ and $\hat{W}_D^{q}(z) = \hat{a}_D^{(q)}(z) / a_D^{(q)}(z)$. Information on $h_1^s(x)$ has been obtained by integrating over $z$ the previous expression in each $x$ bin. The values of $xh_1^s(x)$, as predicted by the Yang model and based on the measured $\Lambda$ polarization, are shown in Figure~\ref{fig:xh1s_y}. An average value has been taken for the scalar and vector diquarks containing or not a strange quark.

%\begin{strip}
%\begin{center}
%  \begin{equation}
%        h_1^s(x) = \frac{1}{\langle\hat{W}_S^{(s)}\rangle} \Bigg \{ P_\Lambda(x) \left[\left(4h_1^u(x)+h_1^d(x)\right)  \cdot \frac{1}{4} \left[\langle F_S^{(u/s)}\rangle +3\langle %F_M^{(u/s)}\rangle\right] +f_1^s(x) \right] - \left(4h_1^u(x)+h_1^d(x)\right)  \cdot \frac{1}{4} \left[\langle \hat{W}_S^{(u)}\rangle \langle F_S^{(u/s)}\rangle -\langle %\hat{W}_V^{(u)}\rangle\langle F_M^{(u/s)}\rangle\right]  \Bigg \}
%\end{equation}
%\end{center}
%\end{strip}

\newpage
\section{Summary}

The transversity-induced polarization has been measured in COMPASS for $\Lambda$ and $\bar{\Lambda}$ hyperons using a transversely polarized proton target and a 160 GeV/$c$ muon beam and found compatible with zero in various kinematic regions. All the COMPASS data, namely all the existing data in the world suitable for this measurement, have been used. Still, the statistical uncertainty on the measured polarization is rather large and it is not possible to draw definite conclusions. Under the hypothesis that transversity is a valence object, the data have been used to investigate the ratio of the $z-$integrated polarized and unpolarized fragmentation function. If instead a non relativistic SU(3) quark model or a quark-diquark model are considered, some information can be derived on the transversity distribution for the $s$ quark.

\vspace{2cm}
%\bibliography{bibfilepos.bib}

\begin{thebibliography}{10}

\bibitem{Baldracchini:1980uq}
F.~Baldracchini, N.~S. Craigie, V.~Roberto and M.~Socolovsky, \emph{{A Survey
  of Polarization Asymmetries Predicted by {QCD}}},
  \href{https://doi.org/10.1002/prop.19810291102}{\emph{Fortsch. Phys.}
  {\bfseries 30} (1981) 505}.

\bibitem{Artru:1990wq}
X.~Artru and M.~Mekhfi, \emph{{What can we learn from unpolarized and polarized
  electroproduction of fast baryons?}},
  \href{https://doi.org/10.1016/0375-9474(91)90709-F}{\emph{Nucl. Phys.}
  {\bfseries A532} (1991) 351}.

\bibitem{Kunne:1993nq}
R.~A. Kunne, J.~Arvieux, P.~A.~M. Guichon, J.~M. Laget, F.~Kunne, B.~Saghai
  et~al., \emph{{Electroproduction of polarized Lambdas (a proposal for the
  European Electron Facility)}}, {\emph{Italian Phys. Soc. Proc.} {\bfseries
  44} (1993) 401}.

\bibitem{Anselmino:2003wu}
M.~Anselmino, \emph{{Transversity and Lambda polarization}},  in
  \emph{{Workshop on Future Physics at COMPASS Geneva, Switzerland, September
  26-27, 2002}}, 2003, \href{https://arxiv.org/abs/hep-ph/0302008}{{\ttfamily
  hep-ph/0302008}}.

\bibitem{Avakian:2016rst}
H.~Avakian, A.~Bressan and M.~Contalbrigo, \emph{{Experimental results on
  TMDs}}, \href{https://doi.org/10.1140/epja/i2016-16150-x,
  10.1140/epja/i2016-16165-3}{\emph{Eur. Phys. J.} {\bfseries A52} (2016) 150}.

\bibitem{Commins:1983ns}
E.~D. Commins and P.~H. Bucksbaum, \emph{{Weak Interactions of Leptons and
  Quarks}}. 1983.

\bibitem{Mulders:1995dh}
P.~J. Mulders and R.~D. Tangerman, \emph{{The complete tree level result up to
  order 1/Q for polarized deep inelastic leptoproduction}},
  \href{https://doi.org/10.1016/S0550-3213(96)00648-7,
  10.1016/0550-3213(95)00632-X}{\emph{Nucl. Phys.} {\bfseries B461} (1996) 197}
  [\href{https://arxiv.org/abs/hep-ph/9510301}{{\ttfamily hep-ph/9510301}}].

\bibitem{Barone:2003fy}
V.~Barone and P.~G. Ratcliffe, \emph{{Transverse spin physics}}. 2003.

\bibitem{Boer:1999uu}
D.~Boer, R.~Jakob and P.~J. Mulders, \emph{{Angular dependences in electroweak
  semiinclusive leptoproduction}},
  \href{https://doi.org/10.1016/S0550-3213(99)00586-6}{\emph{Nucl. Phys.}
  {\bfseries B564} (2000) 471}
  [\href{https://arxiv.org/abs/hep-ph/9907504}{{\ttfamily hep-ph/9907504}}].

\bibitem{Ferrero}
A.~Ferrero~on~behalf~of~the COMPASS~Collaboration, \emph{Measurement of
  transverse {$\Lambda$} and {$\bar{\Lambda}$} polarization at {COMPASS}},
  \href{https://doi.org/10.1063/1.2750815}{\emph{AIP Conference Proceedings}
  {\bfseries 915} (2007) 436}.

\bibitem{Negrini:2009zz}
T.~Negrini~on~behalf~of~the COMPASS~Collaboration, \emph{{Lambda polarization
  with a transversely polarized proton target at the COMPASS experiment}},
  \href{https://doi.org/10.1063/1.3215731}{\emph{AIP Conf. Proc.} {\bfseries
  1149} (2009) 656}.

\bibitem{Armenteros1}
J.~Podolanski and R.~Armenteros, \emph{{III}. analysis of {V}-events},
  \href{https://doi.org/10.1080/14786440108520416}{\emph{The London, Edinburgh,
  and Dublin Philosophical Magazine and Journal of Science} {\bfseries 45}
  (1954) 13}.

\bibitem{Armenteros2}
R.~Armenteros, K.~Barker, C.~Butler, A.~Cachon and C.~York, \emph{{LVI}. {The}
  properties of charged {V}-particles},
  \href{https://doi.org/10.1080/14786440608520216}{\emph{The London, Edinburgh,
  and Dublin Philosophical Magazine and Journal of Science} {\bfseries 43}
  (1952) 597}.

\bibitem{Adolph:2013dhv}
C.~Adolph~et~al. (COMPASS~Collaboration), \emph{{Study of $\Sigma(1385)$ and
  $\Xi(1321)$ hyperon and antihyperon production in deep inelastic muon
  scattering}},
  \href{https://doi.org/10.1140/epjc/s10052-013-2581-9}{\emph{Eur. Phys. J.}
  {\bfseries C73} (2013) 2581}
  [\href{https://arxiv.org/abs/1304.0952}{{\ttfamily 1304.0952}}].

\bibitem{Alekseev:2010rw}
M.~G. Alekseev~et~al. (COMPASS~Collaboration), \emph{{Measurement of the
  Collins and Sivers asymmetries on transversely polarised protons}},
  \href{https://doi.org/10.1016/j.physletb.2010.08.001}{\emph{Phys. Lett.}
  {\bfseries B692} (2010) 240}
  [\href{https://arxiv.org/abs/1005.5609}{{\ttfamily 1005.5609}}].

\bibitem{Adolph:2012sn}
C.~Adolph~et~al. (COMPASS~Collaboration), \emph{{Experimental investigation of
  transverse spin asymmetries in muon-p SIDIS processes: Collins asymmetries}},
  \href{https://doi.org/10.1016/j.physletb.2012.09.055}{\emph{Phys. Lett.}
  {\bfseries B717} (2012) 376}
  [\href{https://arxiv.org/abs/1205.5121}{{\ttfamily 1205.5121}}].

\bibitem{Ageev:2006da}
E.~S. Ageev~et~al. (COMPASS~Collaboration), \emph{{A New measurement of the
  Collins and Sivers asymmetries on a transversely polarised deuteron target}},
  \href{https://doi.org/10.1016/j.nuclphysb.2006.10.027}{\emph{Nucl. Phys.}
  {\bfseries B765} (2007) 31}
  [\href{https://arxiv.org/abs/hep-ex/0610068}{{\ttfamily hep-ex/0610068}}].

\bibitem{Martin:2014wua}
A.~Martin, F.~Bradamante and V.~Barone, \emph{{Extracting the transversity
  distributions from single-hadron and dihadron production}},
  \href{https://doi.org/10.1103/PhysRevD.91.014034}{\emph{Phys. Rev.}
  {\bfseries D91} (2015) 014034}
  [\href{https://arxiv.org/abs/1412.5946}{{\ttfamily 1412.5946}}].

\bibitem{Yang:2002gh}
J.-J. Yang, \emph{{Flavor and spin structure of quark fragmentation functions
  in a diquark model for octet baryons}},
  \href{https://doi.org/10.1103/PhysRevD.65.094035}{\emph{Phys. Rev.}
  {\bfseries D65} (2002) 094035}.

\bibitem{Anselmino:2000ga}
M.~Anselmino, M.~Boglione and F.~Murgia, \emph{{Lambda and anti-Lambda
  polarization in polarized DIS}},
  \href{https://doi.org/10.1016/S0370-2693(00)00455-X}{\emph{Phys. Lett.}
  {\bfseries B481} (2000) 253}
  [\href{https://arxiv.org/abs/hep-ph/0001307}{{\ttfamily hep-ph/0001307}}].

\bibitem{Buckley:2014ana}
A.~Buckley, J.~Ferrando, S.~Lloyd, K.~Nordstr\H{o}m, B.~Page, M.~R\H{u}fenacht
  et~al., \emph{{LHAPDF6: parton density access in the LHC precision era}},
  \href{https://doi.org/10.1140/epjc/s10052-015-3318-8}{\emph{Eur. Phys. J.}
  {\bfseries C75} (2015) 132}
  [\href{https://arxiv.org/abs/1412.7420}{{\ttfamily 1412.7420}}].

\bibitem{Franco:fe18}
F.~Bradamante~on~behalf~of~the COMPASS~Collaboration, \emph{{The future SIDIS
  measurement on transversely polarized deuterons by the COMPASS
  Collaboration}}, {\emph{these Proceedings} }.

\bibitem{Yang:2001sy}
J.-J. Yang, \emph{{q $\rightarrow$ Lambda fragmentation function and nucleon
  transversity distribution in a diquark model}},
  \href{https://doi.org/10.1016/S0375-9474(01)01281-7}{\emph{Nucl. Phys.}
  {\bfseries A699} (2002) 562}
  [\href{https://arxiv.org/abs/hep-ph/0111382}{{\ttfamily hep-ph/0111382}}].

\bibitem{Jakob:1997wg}
R.~Jakob, P.~J. Mulders and J.~Rodrigues, \emph{{Modeling quark distribution
  and fragmentation functions}},
  \href{https://doi.org/10.1016/S0375-9474(97)00588-5}{\emph{Nucl. Phys.}
  {\bfseries A626} (1997) 937}
  [\href{https://arxiv.org/abs/hep-ph/9704335}{{\ttfamily hep-ph/9704335}}].

\end{thebibliography}
\bibliographystyle{jhep.bst}

\providecommand{\href}[2]{#2}\begingroup\raggedright\endgroup

\end{document}